\documentclass[runningheads]{llncs}
\usepackage[T1]{fontenc}
\usepackage{makecell}
\usepackage{graphicx}
\usepackage[dvipsnames]{xcolor}

\usepackage{tabularx}
\usepackage{tipa}

\usepackage[T2A]{fontenc}
\usepackage[utf8]{inputenc}

\begin{document}

\title{Revealing the Hidden Temporal Structure of HubertSoft Embeddings based on the Russian Phonetic Corpus}
\titlerunning{Hidden Temporal Structure of HubertSoft Embeddings}
%
\author{Anastasia Ananeva \inst{1}\orcidID{0009-0006-4008-5100}\and
Anton Tomilov\inst{2}\orcidID{0000-0002-7159-8607)} \and
Marina Volkova\inst{2}\orcidID{0009-0006-6097-5426}}
\authorrunning{A. Ananeva et al.}
%
\institute{ITMO University, Saint Petersbourg, Russia
\email{464995@niuitmo.ru}\thanks{Corresponding author} \\
\and
STC-innovations Ltd., Russia\\
\email{\{tomilov, volkova\}@speechpro.com}}
\maketitle              
\begin{abstract}
Self-supervised learning (SSL) models such as Wav2Vec 2.0 and HuBERT have shown remarkable success in extracting phonetic information from raw audio without labelled data. While prior work has demonstrated that SSL embeddings encode phonetic features at the frame level, it remains unclear whether these models preserve temporal structure, specifically, whether embeddings at phoneme boundaries reflect the identity and order of adjacent phonemes. This study investigates the extent to which boundary-sensitive embeddings from HubertSoft, a soft-clustering variant of HuBERT, encode phoneme transitions. Using the CORPRES Russian speech corpus, we labelled 20 ms embedding windows with triplets of phonemes corresponding to their start, centre, and end segments. A neural network was trained to predict these positions separately, and multiple evaluation metrics, such as ordered, unordered accuracy and a flexible centre accuracy, were used to assess temporal sensitivity. Results show that embeddings extracted at phoneme boundaries capture both phoneme identity and temporal order, with especially high accuracy at segment boundaries. Confusion patterns further suggest that the model encodes articulatory detail and coarticulatory effects. These findings contribute to our understanding of the internal structure of SSL speech representations and their potential for phonological analysis and fine-grained transcription tasks.

\end{abstract}

\section{Introduction}

Self-supervised learning (SSL) models such as Wav2Vec 2.0~\cite{ref_article1} and HuBERT~\cite{ref_article2} have significantly advanced the field of speech representation learning by extracting rich phonetic information without labelled data. {The embeddings received usually contain a large amount of information which is used in deepfake detection~\cite{ref_article12}}. These models have demonstrated impressive performance on tasks including phoneme classification, automatic speech recognition and articulatory feature prediction. However, it remains underexplored how these embeddings encode phoneme boundaries and whether the temporal structure is also encoded.

It has been shown that SSL embeddings encode temporally localized phonetic cues accessible through supervised probing~\cite{ref_article4}. For instance, linear or shallow MLP probes have effectively predicted phonetic features such as place, manner, and voicing at the frame level~\cite{ref_article5,ref_article6}. These works use Wav2Vec 2.0 to confirm that transitions in the embedding space align closely with phoneme boundaries, offering indirect evidence of temporal structure encoding. Other approaches track articulatory feature trajectories or analyse embedding differences at phoneme transitions, further supporting the presence of boundary-sensitive representations~\cite{ref_article3}.

Phonetic features are also reflected in the embeddings received with HuBERT~\cite{ref_article10,ref_article11}. However, these studies focus mainly on some phonetic features encoded in the embeddings. For example, HuBERT embeddings differentiate vowel quality by encoding distinctions in vowel height, backness, and roundedness. They also effectively represent consonantal articulation features such as place and manner, voicing, nasality, and the stop–fricative contrast. Beyond segmental properties, HuBERT embeddings have been found to reflect prosodic features like pitch, stress, and intonation, and are sensitive to phonetic context effects, including coarticulation.

{In addition, there is growing evidence that both self-supervised and supervised models can be effectively used for phoneme boundary detection. Self-supervised models have been shown to capture abrupt spectral or articulatory transitions in their embeddings, which align closely with phoneme boundaries~\cite{ref_article13}. Techniques such as comparing adjacent frame representations or applying peak detection over embedding dissimilarities allow for boundary identification without explicit labels, suggesting that SSL models implicitly encode segmental structure alongside phonetic content. Complementing this, supervised approaches such as those using Connectionist Temporal Classification (CTC) loss have framed the task as aligning sequences of phoneme pairs to speech. This enables models to focus on transition regions and predict boundaries with high temporal precision~\cite{ref_article14}. Together, these lines of work highlight the capacity of learned representations to reflect temporal dynamics in speech.}

Despite this progress, to the best of our knowledge, no prior study has directly tested whether SSL embeddings encode the identity of the phoneme that begins or ends a segment encoded. The existing literature focuses predominantly on general framewise phonetic classification or the analysis of internal embedding structure. 

This gap is especially pertinent given the growing interest in understanding how SSL models internalize sub-segmental linguistic structure. Notably, there is also a lack of probing studies involving HubertSoft~\cite{ref_article8}, a modification of HuBERT which generates soft posterior probabilities over learned acoustic units instead of hard cluster assignments.

The current work aims to fill this gap by evaluating whether boundary-frame embeddings from HubertSoft can be used to predict the phoneme that begins or ends a segment. By focusing on supervised probing of initial and final frames at phoneme boundaries, we offer a novel perspective on the structure encoded within self-supervised speech embeddings and shed light on the representations’ utility for fine-grained phonological tasks.

\section{Material}
The experiments were conducted using the CORPRES~\cite{ref_article7} Russian speech corpus. {The dataset contains recordings of various speakers, annotated on multiple linguistic levels. Phonetic labeling was performed by trained phoneticians, ensuring a high degree of reliability. However, in certain instances, particularly at the boundaries between acoustically similar sounds, such as between two vowels or a vowel and a consonant, the segmentation may be imprecise. In such cases, the boundary is typically placed approximately at the midpoint of the acoustic transition between the sounds~\cite{ref_book1}}. The hand-labelled recordings from eight speakers were utilized {for embedding extraction with the HubertSoft model. For embedding extraction, the HubertSoft model was used with frozen weights, while only the classification heads were trained.} Initially, embeddings were extracted from the recordings and then averaged within phonetic segment boundaries, providing a compact representation of each labelled sound. These averaged embeddings were used in a subset of experiments to establish baseline performance. The symbols for various sounds were also taken from the CORPRES.

\begin{figure}
\includegraphics[width=\textwidth]{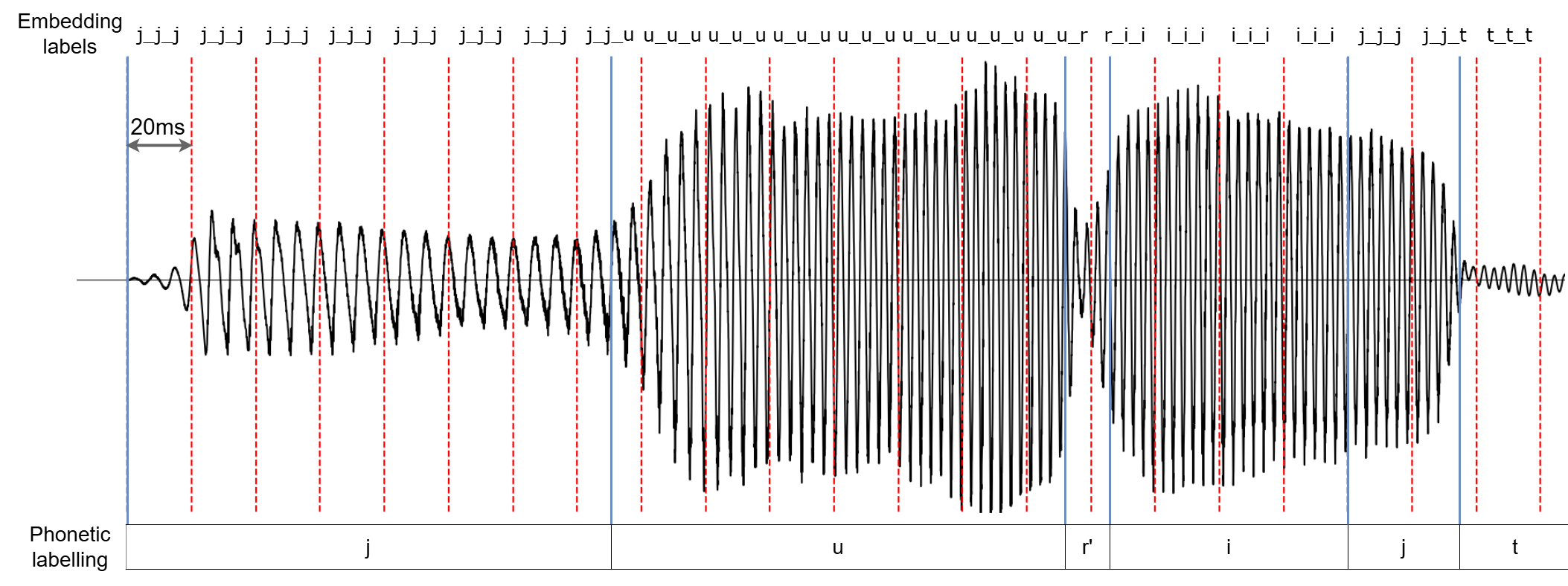}
\caption{{Example of embedding extraction and a triplet label generation based on manual phonetic annotation across 20ms time frames.}} \label{fig1}
\end{figure}

To move towards more fine-grained, automatic analysis suitable for speech transcription tasks, we also extracted non-averaged embeddings. In this setup, each embedding represented a 20 ms window of the audio signal, {as shown in Figure~\ref{fig1},} without overlap. This approach preserved high temporal resolution, which was essential due to the presence of very short phonetic events (as brief as 5 ms) in the corpus. {Such short phoneme realizations are usually unstressed vowels or /r/ and they appear due to the fluency of speech. They may not be pronounced properly, but still are labelled to reflect the phonetic composition of speech.} 

Each embedding was assigned a triplet label indicating the phoneme present at its beginning, centre, and end, derived from the time-aligned phonetic annotations. Table~\ref{tab1} gives some examples of possible labels. This labelling scheme allowed us to capture phonetic transitions within the window. In most cases, all three sub-labels were identical, suggesting that the window was fully inside a single phoneme (typically its central part). However, a substantial number of embeddings spanned phoneme boundaries, where the centre label often matched either the start or end label. These boundary-spanning embeddings are particularly informative for analysing whether the learned representations reflect the temporal structure of the speech signal. 

\begin{table}[ht]
\caption{Examples of labels.}\label{tab1}
\begin{tabularx}{\textwidth}{|>{\centering\arraybackslash}X|>{\centering\arraybackslash}X|>{\centering\arraybackslash}X|}
\hline
Central embedding & Border embedding & Two-border embedding \\
\hline
a\_a\_a, p\_p\_p & a\_p\_p, p\_p\_a, a\_a\_p & a\_p\_a, a\_p\_s \\
\hline
\end{tabularx}
\end{table}

This research focuses mainly on the border embeddings, as they represent a part of a non-homogeneous signal, in which the sequence of sounds matters. Two-border embeddings were excluded as there is a small number of them.

\section{Method}

Our initial experiments demonstrated that HubertSoft embeddings encode substantial phonetically relevant information. In particular, the classification of averaged embeddings (i.e., embeddings averaged over one phoneme) achieved a balanced accuracy of 0.94 on the task of consonant classification. Based on this finding, a similar architecture (a four-layer feed-forward neural network) was adapted for the classification of non-averaged, boundary-sensitive embeddings, with separate output heads corresponding to the start, centre, and end positions of each 20 ms window.

To assess generalization, a speaker-independent evaluation setup was used: recordings from four speakers in the corpus were allocated for training, and the remaining four speakers were used for testing.

For training, only embeddings whose associated labels contained phonemes from a predefined target list were selected, {which allowed us to limit the target group and not to use the whole phoneme inventory of the language}. Each selected embedding was paired with a triplet label marking the phonetic class present at the beginning, centre, and end of the 20 ms window. {To deal with class imbalance the rare labels were avoided and only the sounds which are represented widely enough were chosen.}

The network was trained using the Cross Entropy loss function and the AdamW optimizer. The loss function was computed as the sum of cross-entropy losses across the three output positions. {Some hyper-parameters such as learning rate, the dropout size and weight decay were experimentally adjusted}.


To evaluate whether the HubertSoft embeddings capture the temporal structure of the signal, we trained a neural network to predict phoneme labels corresponding to the beginning, middle, and end parts of a 20 ms segment. Evaluation was conducted using multiple metrics designed to assess different aspects of temporal representation.\textbf{ Ordered accuracy} was computed to measure the proportion of predictions where all three positions (start, centre, end) were correctly classified in the correct sequence, reflecting the model’s ability to recognize and preserve temporal order in the embedding space. To complement this, \textbf{unordered accuracy} was introduced, where predictions were considered correct if they matched the set of target labels regardless of order — testing whether the embedding retains phonetic content without temporal precision.

To better evaluate the model's ability to localize phonetic boundaries while allowing some ambiguity in the center position, we introduce another metric: \textbf{ordered flexible centre accuracy}.

This metric relaxes the strict requirement of full label sequence matching by enforcing correctness at the boundaries while allowing the centre prediction to vary within a constrained range. Specifically, a prediction is considered correct if:
\begin{itemize}
    \item the predicted start and predicted end labels exactly match the corresponding true start and end labels, and
    \item the predicted centre label matches either the true start or true end label.
\end{itemize}

Table~\ref{tab4} summarizes all possible labels considered correct for each metric. Notably, unordered accuracy captures the presence of labels regardless of their order. 

\begin{table}[ht]
\caption{Examples of what is considered to be a correct label for different metrics.}
\label{tab4}
\centering

\begin{tabularx}{\textwidth}{|>{\centering\arraybackslash}X
                               |>{\centering\arraybackslash}m{2.2cm}
                               |>{\centering\arraybackslash}X|}
\hline
\textbf{Metric} & \textbf{True label} & \textbf{Correct predicted label(s)} \\
\hline
Unordered accuracy & a\_p\_p & 
\makecell{a\_p\_p, a\_a\_p, p\_a\_p,\\ p\_p\_a, a\_p\_a, p\_a\_a} \\
\hline
Ordered accuracy & a\_p\_p & a\_p\_p \\
\hline
Ordered flexible centre accuracy & a\_p\_p & a\_p\_p, a\_a\_p \\
\hline
\end{tabularx}
\end{table}
\section{Results}

Classification experiments were conducted to evaluate the model's ability to recognize phonemic content from embeddings. Several phoneme groups were tested, with the most extensive group comprising 6 stressed vowels and 35 consonants, both palatalized and non-palatalized. This group provides a representative sample of the sound inventory and includes diverse acoustic features. The results for this group are presented in Table~\ref{tab2}.

\begin{table}[ht]
\caption{Classification results for the group of vowels and consonants.}
\label{tab2}
\begin{tabularx}{\textwidth}{|>{\centering\arraybackslash}X|>{\centering\arraybackslash}X|}
\hline
\textbf{Metrics} & \textbf{Value} \\
\hline
Ordered accuracy & 0.5312 \\
Unordered accuracy & 0.9069 \\
Ordered flexible centre accuracy & 0.7645 \\
Start accuracy & 0.8823 \\
Centre accuracy & 0.6277 \\
End accuracy & 0.8563 \\
\hline
\end{tabularx}
\end{table}

To establish a baseline for ordered accuracy, we simulated a hypothetical model that correctly identifies which phonemes are present in an embedding (with 0.9 probability per phoneme) but assigns them in a random order. Under these conditions, the expected ordered accuracy is approximately 0.22.

In comparison, our model achieves an ordered accuracy of 0.53 across all phoneme groups. This result significantly exceeds the baseline, indicating the model’s capacity to infer not only phoneme identities but also their temporal arrangement within the segment.

The unordered accuracy, which evaluates phoneme presence regardless of position, is approximately 0.91 for this group. The clear gap between unordered and ordered accuracies suggests that the model captures some degree of temporal structure inherent in the embeddings.

To further investigate positional sensitivity, we analysed classification accuracy for each of the three phoneme positions: start, centre, and end. The central position revealed a noticeable drop in value (0.63), suggesting that the model frequently confuses centre phonemes with those in the start or end positions. In contrast, the start and end positions yielded higher accuracies (0.88 and 0.86, respectively), likely due to more distinct acoustic transitions at the segment boundaries.

The ordered flexible centre accuracy metric, which tolerates minor shifts around the central position, reaches 0.77. This result supports the hypothesis that embeddings preserve temporal phoneme structure to a meaningful extent.
\subsection{Performance Across Sound Groups}
 A consistent pattern emerged while comparing performance across different sound groups, such as vowels, voiced and devoiced plosives, fricatives, palatalized and non-palatalized consonants. It was important to include vowels in the groups of consonants, as these borders allow to receive more examples for every type of sound. 
 
\begin{figure}
\centering
\includegraphics[width=0.6\textwidth]{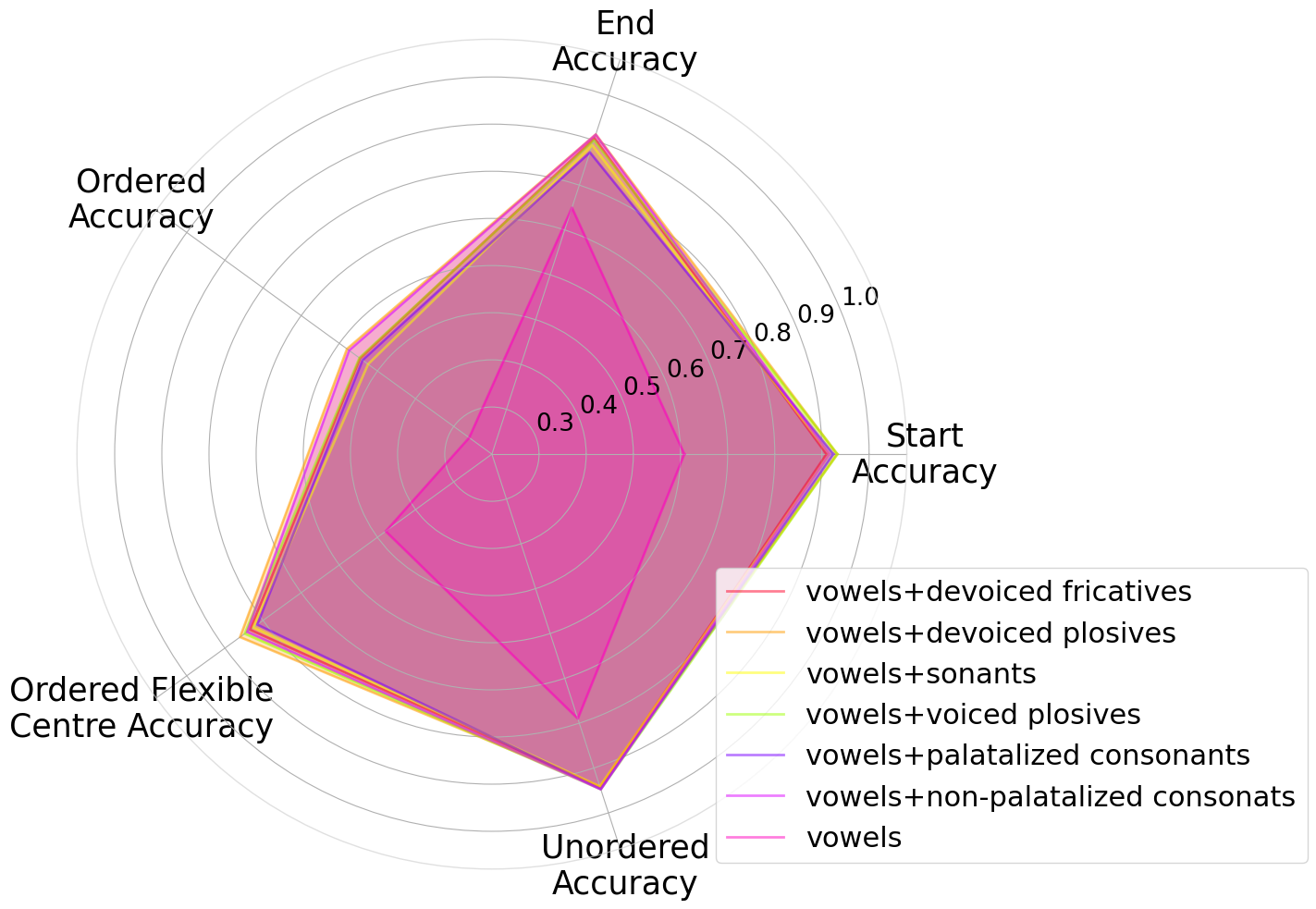}
\caption{{Comparison of classification accuracies across sound types.}} \label{fig}
\end{figure}

 All metrics tended to be highest for the group of voiceless plosives and lowest for sonants and vowels as it is shown in Figure~\ref{fig}. The group of various vowels, both stressed and not stressed, showed the lowest results. This can be attributed to the presence of clear transitions between segments with periodic structure (e.g., vowels) and aperiodic segments (e.g., plosives), which creates sharper acoustic boundaries and facilitates detection.

\begin{figure}
\includegraphics[width=\textwidth]{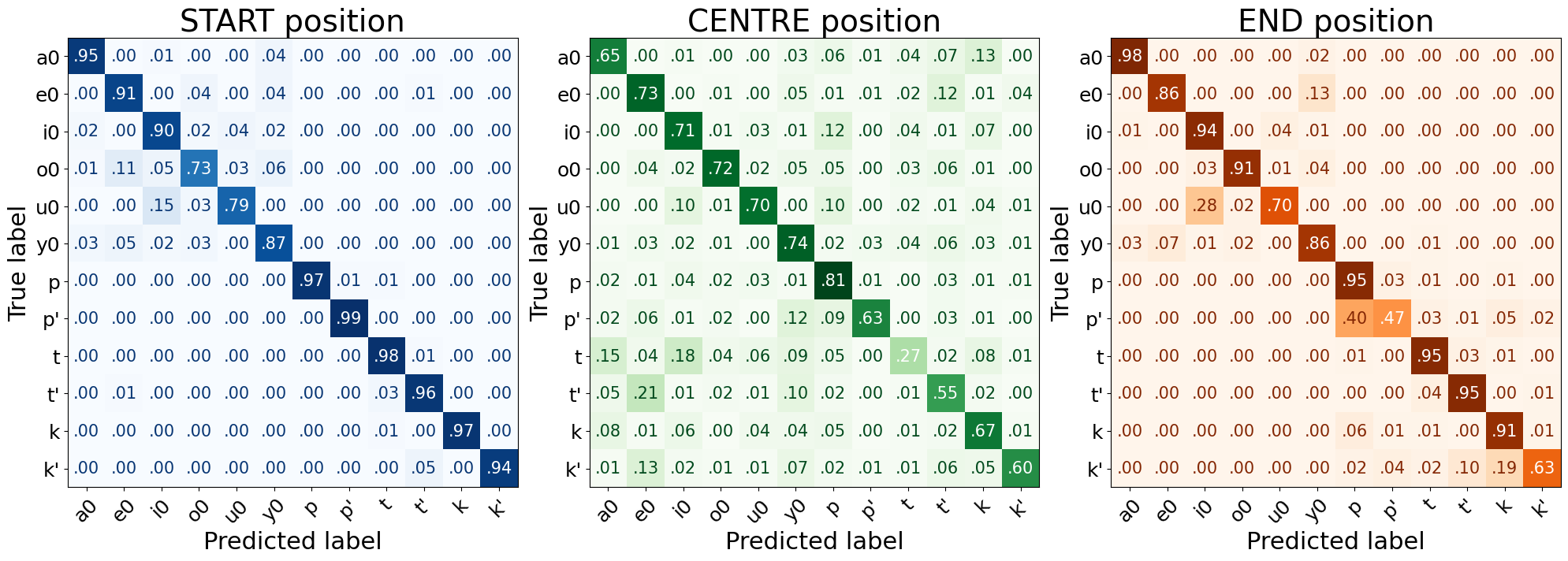}
\caption{Confusion matrices for start, center and end position for the group of plosives and vowels.} \label{fig2}
\end{figure}

A more detailed examination of end-position predictions, particularly for vowels and devoiced plosives, highlights some systematic patterns. For instance, the palatalized plosive labeled as \textit{p'} is frequently confused with its hard counterpart \textit{p} as can be seen in Figure\footnote{The symbols for sounds in figures are taken from the CORPRES labelling.}~\ref{fig2}. In Russian, /p\textsuperscript{j}/ and /p/ are considered a phonemic pair, and this confusion is phonetically plausible: both begin with a closure, and their primary distinction lies in the release phase or even the onset of the following vowel. Since the model classifies the final phoneme in the sequence (which is the onset of the next segment) the embedding appears to encode primarily the closure information, making \textit{p'} and \textit{p} difficult to distinguish at this point.

A similar trend is observed for the \textit{k-k'} pair. Interestingly, the \textit{t-t'} contrast does not exhibit the same level of confusion. Although /t\textsuperscript{j}/ is categorized as a plosive in Russian, acoustically it tends toward a fricative-like realization, often approximating [\t{ts}\textsuperscript{j}]~\cite{ref_article15}. This produces a noisier closure phase, which helps the model differentiate it even at the start of the segment. Thus, acoustic differences rooted in allophonic variation seem to impact the model's ability to distinguish palatalized consonants.


Another example of differentiation between the beginning and the end of can be found in the comparison of the affricate \textit{\t{ts}} to plosive \textit{t} and fricative \textit{s}, which have a common place of articulation.
The results showed the high similarity of \textit{\t{ts}} sound to the \textit{s} sound in the start position, which corresponds to the ending of the first sound. In contrast, it becomes closer to the \textit{t} sound at its beginning, because both have a closure. This shows us that the model classifies the parts of the embedding independently and is definitely capable of distinguishing between ways of articulation.

\begin{figure}
\includegraphics[width=\textwidth]{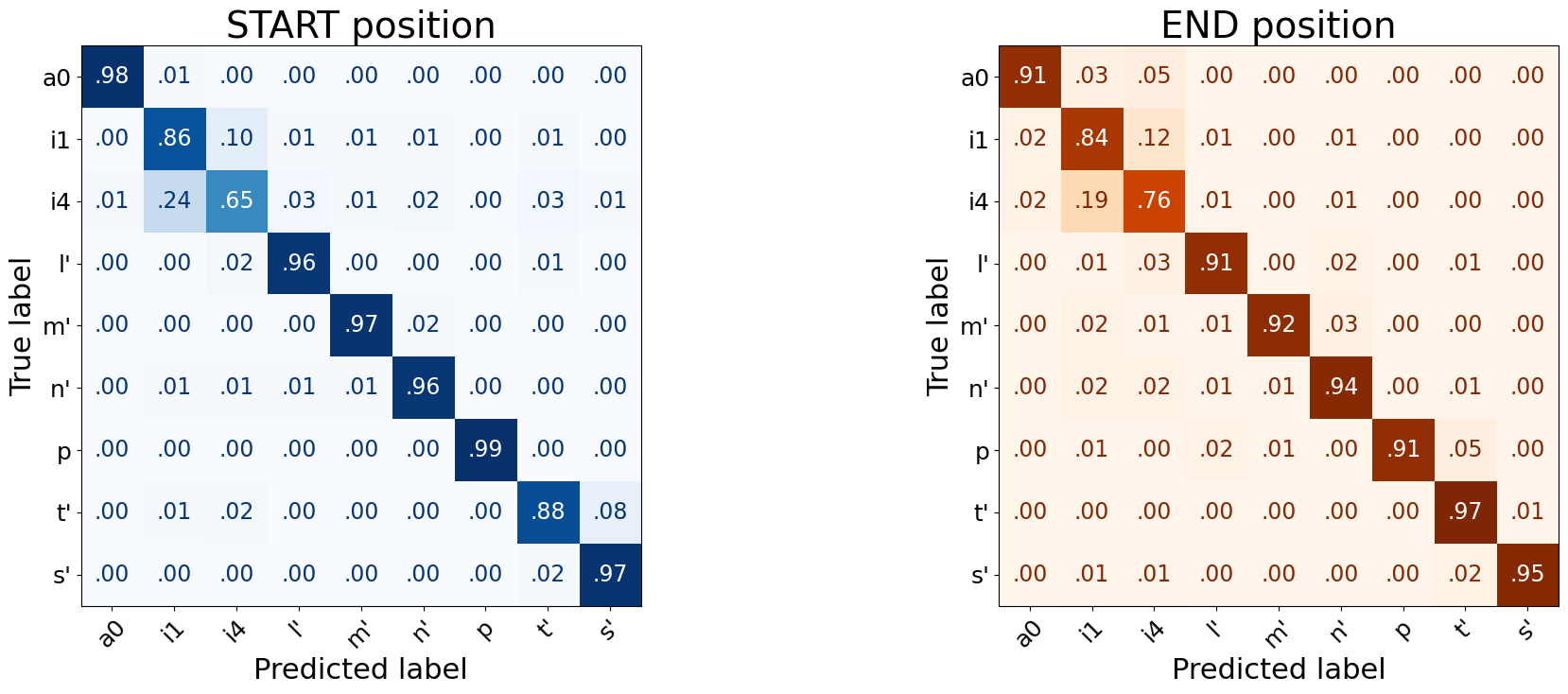}
\caption{Confusion matrices for start, center and end position for the group of palatalized consonants and a0.} \label{fig5}
\end{figure}

A comparable pattern of context-dependent variation is observed in the embeddings for the vowel \textit{a}, particularly in the end position after palatalized consonants. The confusion matrices in Figure~\ref{fig5} show that while \textit{a} (a0 in this case\footnote{0 denotes a stressed vowel position.}) is recognized with high confidence at the start of the segment (0.98), its representation becomes more ambiguous at the end (0.91). Specifically, although the diagonal value for \textit{a0} remains relatively high, there is a noticeable increase in off-diagonal activations toward other vowel classes, \textit{i1}\footnote{1 denotes a pre-stressed vowel position.} and \textit{i0} specifically. This aligns with known coarticulatory effects in Russian, where a preceding palatalized consonant can front the articulation of the following vowel which effectively shifts \textit{a} toward an [i]-like quality~\cite{ref_thesis1}. Since the end-position predictions target the onset of the following segment, the model appears to encode this transitional state in the vowel embedding. The fact that this shift is not evident in the start-position matrix further supports the interpretation that the embedding space captures dynamic phonetic interactions rather than static phoneme identity alone. Thus, the model’s behaviour provides indirect evidence of temporal encoding of coarticulatory influence, particularly for vowel segments in soft-consonant contexts.

\subsection{Application of the Model}
The presence of temporal structure in the learned embeddings enables precise phonetic labelling within short audio segments. {Importantly, this enhancement in temporal resolution was achieved without full retraining of the HuBERT model, highlighting the efficiency of the approach}. To explore this potential, we applied our model, trained using embeddings from the central and boundary regions of segments, to a short recording of approximately 350~ms in length. This recording contained three consecutive sounds: \textit{i1, l,} and \textit{a0}.

It is important to note that the model was pre-trained exclusively on stressed vowels. Therefore, it cannot reliably distinguish between different stress positions, such as \textit{i0} and \textit{i1}, which limits its ability to differentiate between pre-stressed and stressed vowel variants.

\begin{figure}[h]
\includegraphics[width=\textwidth]{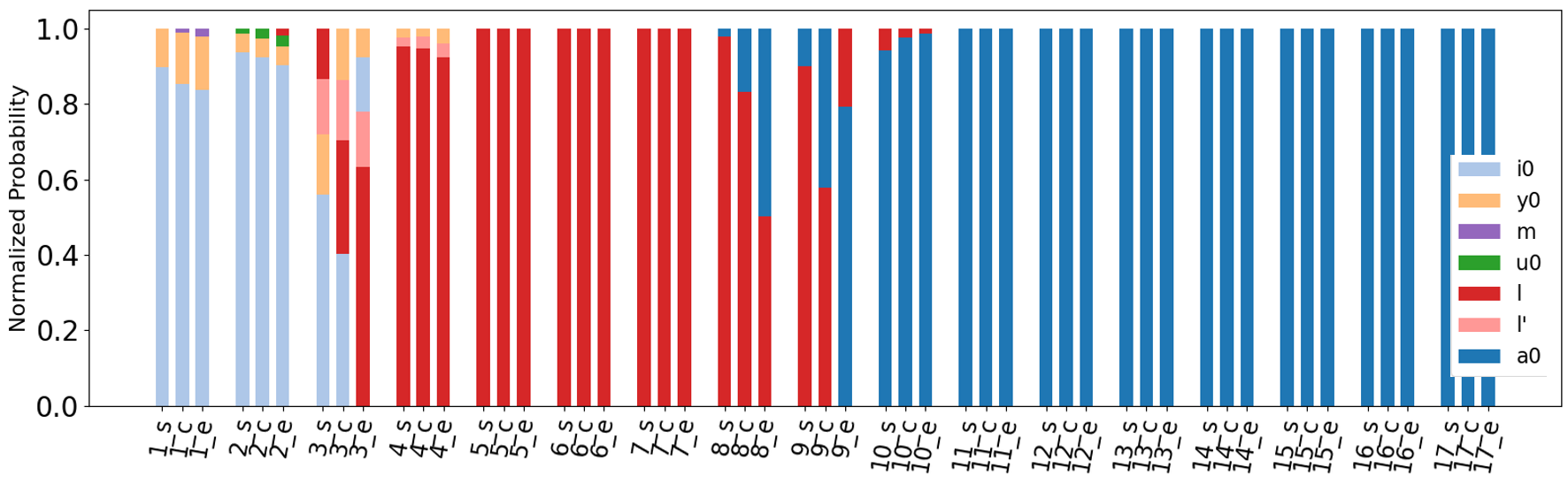}
\caption{The distribution of predicted probabilities for target start(s), center(c) and end(e) sounds across 17 frames for each phoneme position in the sound sequence "i1 l a0".} \label{fig4}
\end{figure}

The model achieved an ordered accuracy of 0.88 when labelling the segment, indicating strong alignment with the expected phoneme sequence. Figure~\ref{fig4} presents the probability distribution across all three phoneme positions over 17 frames. A clear boundary between the first and second sound is visible around frame 3, while frames 7 and 8 capture the transition from \textit{l} to \textit{a}, corresponding to the end of \textit{l} and start of \textit{a}, respectively.

In fact, the only mistake that was achieved is that the border between \textit{i0} and \textit{l} should have been in the frame number 4 according to the initial labelling. The smoothness of transition between a vowel and a sonant makes it impossible to define the precise location of a boundary.

This example illustrates the model’s effectiveness in capturing temporal dynamics within embeddings and demonstrates the feasibility of using such representations for fine-grained phonetic segmentation and labelling.

\section{Discussion and Conclusion}

This study explored the extent to which self-supervised speech embeddings, produced by HubertSoft, encode information related to segment boundaries and temporal structure. The results demonstrate that they not only retain phoneme identity but also encode temporal order with a level of precision sufficient for boundary-aware classification. The model achieved high unordered accuracy, indicating that phonetic content is consistently preserved. Crucially, its ordered and flexible-centre accuracies significantly outperformed chance baselines, suggesting a strong sensitivity to phoneme positioning within the temporal window.

Detailed analysis across phoneme groups revealed that boundary encoding is particularly robust for consonants with distinct acoustic transitions, such as voiceless plosives. In contrast, vowels and sonorants showed lower accuracy, likely due to their smoother spectral characteristics and higher susceptibility to coarticulation. Moreover, some phonetic effects found on the boundaries are also reflected in the embeddings.

These findings show that self-supervised embeddings do not merely encode static phonetic labels but also reflect temporal and contextual dependencies relevant for linguistic analysis. Moreover, the proposed triplet-label probing setup offers a new way to assess the temporal encoding capacity of SSL models.

Summing up, HubertSoft embeddings prove capable of encoding phonemic timing and boundary information, reinforcing the utility for downstream phonological tasks and motivating further work into context-aware, temporally structured speech representations. In future studies, the proposed approach may be applied to explore the internal structure of speech sounds in more detail, potentially contributing to more precise and linguistically informed phonetic labelling.

\begin{credits}

\subsubsection{\discintname}
The authors have no competing interests to declare that are relevant to the content of this article.
\end{credits}
%
%
%
%

\end{document}